\documentclass{elsart}
\newcommand{\beq}{\begin{equation}}
\newcommand{\eeq}{\end{equation}}
\newcommand{\bea}{\begin{eqnarray}}
\newcommand{\eea}{\end{eqnarray}}
\newcommand{\du}{d_U}
\newcommand{\dbz}{d_{BZ}}

\newcommand{\lagr}{{\mathcal L}}
\newcommand{\opr}{{\mathcal O}}
\newcommand{\unp}{U}

\usepackage{graphicx}
 \usepackage{setspace}
 \usepackage{amsfonts}
 \usepackage{amsmath}  
 \usepackage{txfonts} 
\usepackage[]{epsfig}
\begin{document}

\begin{frontmatter}

\title{Vector unparticle enhanced black holes: exact solutions and thermodynamics}

\author{J. R. Mureika\thanksref{jrm}}
\thanks[jrm]{e-mail address: jmureika@lmu.edu}
\address{Department of Physics, Loyola Marymount University, Los Angeles, CA 90045-2659}

\author{Euro Spallucci\thanksref{infn}}
\thanks[infn]{e-mail address: spallucci@ts.infn.it }
\address{Dipartimento di Fisica Teorica, Universit\`a di Trieste
and INFN, Sezione di Trieste, Italy}

\begin{abstract}
Tensor and scalar unparticle couplings to matter have been shown to enhance gravitational 
interactions and provide corrections to the Schwarzschild metric and associated black hole 
structure.  We derive an exact solution to the Einstein equations for vector unparticles, 
and conclusively demonstrate that these induce Riessner-Nordstr\"om (RN)-like solutions 
where the role of the ``charge'' is defined by a composite of unparticle phase space parameters. 
These black holes admit double-horizon structure, although unlike the RN metric these solutions 
have a minimum inner horizon value.  In the extremal limit, the Hawking temperature is shown to vanish.  
As with the scalar/tensor case, the (outer) horizon is shown via entropy considerations to behave like a
fractal surface of spectral dimension $d_H = 2\du$.
\end{abstract}

\end{frontmatter}

\section{Introduction}

Recently, it was proposed that there could be a conformally scale-invariant particle sector of unknown composition 
with a non-trivial IR fixed point \cite{Georgi:2007ek,Georgi:2007si}, at which stronger couplings to the standard 
model emerge.  Dubbed ``unparticle physics'' because of the non-intuitive phase space structure, its introduction 
has caused a flurry of research into modifications to known physics and post-TeV predictions.   
Accelerator phenomenology has been the main emphasis in the literature 
\cite{Liao:2007ic,Liao:2007fv,Liao:2007bx,Rizzo:2007xr,Cheung:2007ap,Bander:2007nd,Cheung:2007jb,Cheung:2007zza}
\cite{Kikuchi:2007qd,Kikuchi:2008pr}, 
but astrophysical/cosmological 
\cite{Freitas:2007ip,Bertolami:2009qn,Hannestad:2007ys,Davoudiasl:2007jr,Das:2007nu,Alberghi:2007vc}
\cite{McDonald:2007bt,Lewis:2007ss,Chen:2007qc,Kikuchi:2007az,Bertolami:2009jq,dejan1}
\cite{Das:2007cc,Deshpande:2007mf,Goldberg:2007tt}, 
low to ultra-high neutrino phenomenology \cite{nu1,nu2,nu3,nu4,nu5,nu6,nu7,nu8} and general quantum field theory 
\cite{Stephanov:2007ry,Kazakov:2007fn,Kazakov:2007su,Lee:2007xd,Gaete:2008aj,Gaete:2008wg}
\cite{Nicolini:2010nb,Nicolini:2010bj} 
applications have been of extreme importance as well. 
 
Constraints on the unparticle parameters $\Lambda_\unp$, the BZ messenger mass $M_\unp$, and the unparticle 
and Banks-Zaks dimensions $\du, d_{BZ}$ are obtained through limits on measurable accelerator phenomenology, 
astrophysical and cosmological observations.  The aforementioned parameters serve to fix the energy and distance 
scale at which the interactions become relevant.  It has been shown that, if $\Lambda_\unp \sim 1~$TeV, then 
strict limits may be placed on the messenger $M_\unp$ for various values of the unparticle dimension $\du$.  
Higher values of $\du$ imply lower values of $M_\unp$, whose value could be as small as a few hundred TeV.
 
One of the most intriguing aspects of unparticle physics is that interactions with standard model particles 
effectively modify the usual gravitational coupling strength \cite{Deshpande:2007mf,Goldberg:2007tt,Mureika:2007nc}.  
Dubbed ``ungravity'', in the Newtonian limit this is most likely to be observed as deviations in planetary orbits 
and perihelion precession \cite{Das:2007cc,Deshpande:2007mf} on large scales, as well as constrain 
Big Bang Nucleosynthesis \cite{Bertolami:2009qn}, dark energy \cite{dejan1} and even entropic gravity \cite{Nicolini:2010nb}.  
Conversely, the very small-scale behavior of scalar and tensor ungravity begins to mimic that of $n$ large 
extra compactified dimensions \cite{ArkaniHamed:1998nn}, but with $n\rightarrow 2\du-2$ and the Newtonian potential
proportional to $1/r^{2\du-1}$  \cite{Mureika:2007nc,Mureika:2008dx,Lee:2009dda}.  
The interesting property here is that, since $\du$ can be non-integer, ungravity reproduces the phenomenology not 
only of standard extra-dimensional physics, but also of a ``fractal'' spacetime.

It was conjectured by perturbative arguments that such a modification to the Newtonian potential would result in 
unparticle-driven mini-black hole creation in high energy collisions \cite{Mureika:2007nc,Mureika:2008dx}.  
More recently, exact solutions to Einstein's equations were derived for unparticle interactions with matter, 
showing that the previous approximation holds in both the weak and strong gravity limits \cite{Gaete:2010sp}.

This paper will address the influence of vector unparticle interactions with matter, and the respective solutions  
of the Einstein equations.  We show that, as in the scalar/tensor case, vector unparticles modify the metric in an 
analogous fashion and admit black hole solutions enhanced by the unparticle parameters.  Since vector ungravity is 
repulsive, however, the resulting horizon and singularity structure is comparable to Riessner-Nordstr\"om class of 
metrics, where the ``charge'' is a composite of unparticle parameters.  We also discuss the unique thermodynamics of such 
black holes, and consider the associated implications for the spacetime dimensionality..

\section{Basics of unparticle physics}
Unparticle physics is characterized by its non-integer scaling dimension $\du$ in phase space, 
making it ``look like'' a system of $\du$ fundamental particles.  A weakly-coupled Banks-Zaks (BZ) 
field \cite{Banks:1981nn} exchanges a massive particle $M_{\unp}$ with standard model field,  suppressed by 
non-renormalizable interactions
\beq
{\mathcal L} = \frac{1}{M^{d_{SM} + d_{BZ} - 4}_{\unp}} \opr_{SM}\opr_{BZ}~~.
\eeq  Here, $\opr$ is the unparticle operator, which may possess any Lorentz type (scalar, vector, tensor, spinor).  
The dimensions $d_{SM}$ and $d_{BZ}$ correspond to the standard model and Banks-Zaks fields. 

The coupling $M_\unp$ will run below some energy scale $\Lambda_\unp < M_\unp$, and the field transmutes to the 
unparticle operator $\opr_\unp$ of dimension $\du \ne \d_{BZ}$.  In this limit, the interaction is
\beq
\lagr = \frac{\kappa}{\Lambda^{k_\unp}} \opr_{SM} \opr_\unp
%~~,~~\kappa = C_\unp \left(\frac{\Lambda_\unp}{M_{BZ}}\right)^k~~,
\eeq 
with $k_\unp = d_{SM}+d_\unp-4$ and $\kappa$ redefined accordingly so the action is dimensionless.   
Since unparticle interactions are heretofore undiscovered, the lower-limit on the energy scale must 
be $\Lambda_\unp \geq 1~$TeV, making it an ideal framework for high energy phenomenology.  

In an attempt to provide a concrete physical mechanism for such a non-physical phase space, several explanations 
have been put forth as to the nature of unparticle stuff.  These include a composite Banks-Zaks particle with a 
continuum of masses \cite{kraz,nikolic,mcdonald2}, or alternatively a Sommerfeld-like model of massless fermions 
coupled to a massive vector field \cite{georgi2}.  Recently, it was also shown that unparticle-like propagators 
may be mimicked by a small collection of ordinary particles via Pad\'e approximations \cite{mpv}.

Vector-like unparticle operators $\opr_\unp$ couple to baryon currents with dimensionless strength $\lambda$ 
according to the interaction
\beq
\lagr= \frac{\lambda}{\Lambda_\unp^{\du-1}} B_\mu \opr^\mu_\unp~~,
\eeq
which will yield an effective potential of the form \cite{Deshpande:2007mf}
\beq
V_\unp(r) \sim \frac{\lambda B_1 B_2}{r^{2\du-1}} \longrightarrow \frac{\lambda m_1 m_2}{M_B^2 r^{2\du-1}}
\eeq
where the baryon numbers for the interacting masses are $B_j \approx m_j/M_B$, and $M_B$ is the baryon mass.  
The modified gravitational potential is then
\beq
\Phi(r) = \Phi_N(r) \left[ 1 - \frac{1}{2\pi^{2\du}}\frac{\Gamma(\du+\frac{1}{2})
\Gamma(\du-\frac{1}{2})}{\Gamma(2\du)} \left(\frac{R_{*v}}{r}\right)^{2\du-2} \right] = 
\Phi_N(r) \left[ 1- \left(\frac{R_{v}}{r}\right)^{2\du-2}\right]
\label{vecV}
\eeq
with the new length scale $R_{v}$ dependent on the coupling strength $\lambda$ and the other unparticle 
parameters.  Equation~\ref{vecV} highlights the repulsive nature of vector ungravity, which as we will see is 
crucial in determining the unique properties of the associated black hole solutions in the relativistic theory.

\section{Vector unparticles corrections to the metric}
The physical system we are going to investigate is an ``hybrid'' of classical matter, classical 
gravity, and ``quantum'' un-gravity due to the exchange of un-vectors.  An initial treatment of
the problem has been done in the weak-field, perturbative regime \cite{Mureika:2007nc,Mureika:2009ii}, but 
here we present a robust derivation from first principles.  The following derivation assumes
$\dbz \approx 1$, but departures from this value are considered more extensively in \cite{Mureika:2009ii}.  

The action for this system is the sum of a classical functional $S_M$ for matter, and a \textit{non-local} 
effective action $S_U$ smoothly extending the Einstein-Hilbert action to include un-vectors dynamics,

\begin{equation}
 S\equiv S_M + S_U
\end{equation}
$S_M$ is the classical matter action for a massive, point-like, particle ``sitting'' in the origin.
 There is some freedom to choose the  explicit form of this functional. Simplicity suggests to 
 introduce $S_M$ in the form of the action for pressure-less, static fluid,  with a ``singular'' 
 (but integrable!) energy density mimicking a ``point-mass'':
\begin{equation}
 S_M\equiv -\int d^4x \sqrt{g}\,\rho\left(\, x\,\right)\, u^\mu\, u^\nu\ ,\quad 
 \rho\left(\, x\,\right)\equiv
 \frac{M}{\sqrt{g}}\int d\tau \,\delta\left(\, x -x\left(\tau\right)\,\right)
\end{equation}

The un-gravity action is obtained by combining the Einstein-Hilbert functional and the non-local
effective action obtained in \cite{Gaete:2008wg} :
\begin{equation}
S_U = \frac{1}{2\kappa^2}\,\int d^4x \sqrt{g}\,\left[ \,
1+\frac{ A_{d_U}}{\left(\,2d_{U}-1\,\right)\sin\left(\,\pi\, d_U\,\right)}
\frac{\kappa_\ast^2}{\kappa^2}
\left(\, \frac{-D^2}{\Lambda^2_U}\,\right)^{1-d_U}\, \right]^{-1} R
\label{ueh}
\end{equation}

where, $D^2$ is the generally covariant D'Alembertian, which can be treated in the Schwinger representation
\begin{equation}
\left(\, D^2\,\right)^{d_U-1}=\frac{1}{\Gamma\left(\, 1-d_U\,\right)} \int_0^\infty ds s^{ -d_U} e^{-s D^2}
\ ,\qquad d_U>1
\nonumber
\end{equation}
The coefficient in the numerator of the correction is 
\begin{equation}
A_{d_U}\equiv \frac{16\pi^{5/2}}{\left(\, 2\pi\,\right)^{2d_U}}
\frac{\Gamma\left(\,d_U + 1/2\,\right)}{ \Gamma\left(\,d_U - 1\,\right)
\Gamma\left(\,2d_U \,\right)}
 \end{equation}
and $\kappa_\ast$ is the coupling between gravity and un-particle. In the vector
case

\begin{equation}
\kappa_\ast \equiv -\frac{\pi}{M_{Pl.}}\left(\,
\frac{\lambda\, M_{Pl.} }{M_B}\,\right)^2
\label{unv}
\end{equation}

where, $M_B\sim 1\, GeV$ is the baryon mass.
Notice the minus sign taking into account the repulsive nature of the interaction.\\
 As the form of the effective action (\ref{ueh}) holds for any kind of unparticle,
 let us proceed without specifying the coupling constant $\kappa_\ast$, and insert
 eq.(\ref{unv}) only in the final result.\\
Our main
purpose is to solve the field equations derived from $S$ by assuming
the source is static, i.e. the four-velocity field $u^\mu$ has only 
non-vanishing time-like component

\begin{equation}
u^\mu\equiv \left(\, u^0\ , \vec{0} \,\right)\ ,\quad
u^0=\frac{1}{\sqrt{-g^{00}}}
\end{equation}

Einstein equations are obtained by varying the action (\ref{ueh})
with respect to the metric $g_{\mu\nu}$. By neglecting surface terms
coming from the variation of the generally covariant D'Alembertian, we find

\begin{eqnarray}
 R^\mu_\nu-\frac{1}{2}\delta^\mu_\nu\, R &&= \kappa^2\,
 \left[ \, 1+
 \frac{A_{d_U}\Lambda^{2-2d_U}_U}{\left(\, 2d_U-1\,\right)\,\sin\left(\,\pi\, d_U\,\right) } 
 \frac{\kappa_\ast^2 }{\kappa^2}
\left(\,-D\,\right)^{d_U-1}\,\right] \,
T^\mu{}_\nu\nonumber\\
&&\equiv \kappa^2\, T^\mu{}_\nu +\kappa_\ast^2 \frac{A_{d_U}}{ \sin\left(\,\pi\, d_U\,\right) }
T_U{}^\mu_\nu
\label{e1}
\end{eqnarray}

In Eq.(\ref{e1}) we have ``shifted'' the un-particle terms to the 
 r.h.s. leaving the l.h.s. in the canonical form. As a matter of fact,
Eq. (\ref{e1}) can be seen as `ordinary'' gravity coupled to an
``exotic'' source term, instead of un-gravity produced by an ordinary
particle. The two interpretations are physically equivalent.\\
The  energy-momentum tensor $T^\mu{}_\nu$ is given by \cite{DeBenedictis:2007bm}

\begin{eqnarray}
&& T^0_0=-\frac{M}{4\pi\, r^2}\delta\left(\, r\,\right) \label{t00}\\
&& T^r_r=0\\
&& T^\theta_\theta=T^\phi_\phi= -\frac{M}{16\pi\, r}\delta\left(\, r\,\right)
\frac{1}{g_{00}}\partial_r\, g_{00}
\end{eqnarray}

where, $T^\theta_\theta\ ,T^\phi_\phi$ are determined by the requirement
$\nabla_\mu T^{\mu\nu}=0$.\\
With this kind of energy-momentum tensor the $00$ and $rr$ components
of the metric tensor turn out to be of the form
\begin{equation}
g_{rr}^{-1}= 1- \frac{2}{r}\, M\left(\, r\,\right)=-\frac{e^{-h_0}}{g_{00}} 
\end{equation}

where the constant $h_0$ can be freely re-absorbed into the deviation of
the time coordinate, and

\begin{equation}
M(r)=-4\pi \int_r^\infty dr \; r^2 T^0_0\  ,\qquad r>0
\label{linm}
%M\left(\, r\,\right)\equiv 4\pi\,\int dr\, r^2 \, T^0_0 \label{linm}
\end{equation}

In Equation (\ref{linm}) the symbol $\int dr$ indicates an indefinite 
integration. 
The constant factor $e^{h_0}$ can be safely rescaled to $1$ by
a redefinition of the time coordinate.\\

We find,

\begin{equation}
M\left(\, r\,\right)=
\frac{2^{2d_U-2}}{4\pi^{1/2}} 
\frac{\Gamma\left(\,d_U -1/2\,\right)}{\Gamma\left(\,2-d_U \,\right) }
\, M \Lambda^{2-2d_U}_U\,\left(\,\frac{1}{r}\,\right)^{2d_U-2}
\end{equation}

and

\begin{eqnarray}
g_{rr}^{-1}&&= -g_{00}= 
=1+ V_N\left(\, r\,\right)\left[\, 1- \left(\,\frac{R_v}{r}\,\right)^{2d_U-2}\,\right]
\label{unschw}\\
R_v &&\equiv \left[\, \frac{1}{2\pi^{2d_U}}
\frac{\Gamma\left(\,d_U -1/2\,\right)\Gamma\left(\,d_U +1/2\,\right)}
{\Gamma\left(\,2d_U\,\right)}\,\right]^{\frac{1}{2d_U -2}}\left(\,
\frac{\lambda\, M_{Pl.} }{M_B}\,\right)^{\frac{1}{d_U-1}}\, \Lambda_U^{-1}
\end{eqnarray}

where, $R_s=2MG_N=2M/M^2_{Pl.}$ is the Schwarzschild radius;
$V_N\left(\, r\,\right)$ is the Newton gravitational potential, and $R_v$
is the new gravitational length scale.

The horizon curve is obtained by the condition $g_{rr}^{-1}(r_H)=0$ 

\begin{equation}
 M=\frac{r_H}{2}\frac{1}{1 -\left(\,R_v /r_H\,\right)^{2d_U-2}}\equiv M\left(\, r_H\,\right)
\label{horizon}
\end{equation}

The intersections between the line $M=\mathrm{const}$ and the curve $M\left(\, r_H\,\right) $
gives the radii of the inner and outer horizons.  In this regard, we notice a first
difference with respect the RN metric, where the inner horizon, $r_-$, can be arbitrarily
small. As the mass $M$ is positive definite, we see from Eq.(\ref{horizon}) that $r_H> R_v$.
That means that the whole horizon curve is shifted to the right by an amount equal to $R_v$.
Thus, $r_-$ can never be smaller than $R_v$.\\
If we decrease $M$ the two horizons approach one to the other and finally will merge
into the single degenerate horizon of an \textit{extremal black hole}. The mass and the
radius of the extremal configuration can be obtained from Eq.(\ref{horizon}) and the
condition

\begin{equation}
 \left(\, \frac{dM}{dr}\,\right)_{r=r_e}=0 
\end{equation}
 
Thus, we find

\begin{eqnarray}
&& r_e=\left(\, 2d_U -1\,\right)^{\frac{1}{2d_U-2}}\, R_v\ ,\label{rextr} \\
&& M_e=\frac{\left(\, 2d_U-1\,\right)^{\frac{2d_U-1}{2d_U-2}}}{4\left(\,d_U-1 \,\right)}\, R_v\ ,
\quad d_U> 1 \label{mextr}
\end{eqnarray}
This result allows us to distinguish three different cases:
\begin{enumerate}
 \item $M> M_e$ Massive objects. They are  two-horizons black holes 
 \item $M= M_e$ Critical objects. They are extremal black hole with a single degenerate horizon
 \item  $M< M_e$ Light objects. They would be ``naked-singularity'' , where no horizon
 shields the curvature singularity in $r=0$.
\end{enumerate}
$M_e$ represents the lower bound for the mass of a vector unparticle modified black hole.
As we shall see in the next section, the extremal black hole has vanishing Hawking
temperature and represents an asymptotic final stage of the evaporation process.

The conditional tense is necessary in the case of light objects, as they have not to be taken too seriously. 
Indeed, the appearance of a naked-singularity is an alarm signal
that the theory we are using is blowing up, rather than a legitimate physical effect. 
 Indeed, invocation of the Cosmic Censorship Conjecture negates the formation of
such black holes, and can in fact be used to constrain the unparticle phase space in
this situation \cite{Mureika:2009ii}.

Divergence in the Riemann curvature, or tidal forces, at the origin is  
the unavoidable side-effect of modeling the source of the field as a ``point-mass''.
By packing a finite energy inside a vanishing spacelike volume disrupts  
the spacetime fabric itself. This is not a physical effect,  but it is the due response
of a classical theory, i.e. General Relativity, to an unphysical infinite density source.
At short distance from the origin General Relativity must be supplemented by
Quantum Mechanics inputs in order to provide self-consistent results
\cite{Nicolini:2005vd,Spallucci:2006zj,Ansoldi:2006vg,Spallucci:2008ez,Nicolini:2009gw}. 
The whole model we are discussing here, can be trusted only far away from the
Planck scale, where, not only matter, but gravity itself must be upgraded to some
proper quantum theory.

\section{Thermodynamics}
Scalar and tensor unparticle-enhanced black hole thermodynamics and their evaporation modes have been addressed previously \cite{Mureika:2008dx,Gaete:2010sp,dejan2}.  In the case of vector ungravity, the Hawking temperature is

\begin{equation}
 T_{d_U}=
 \frac{1}{ 4\pi\, r_+\left[\, 1 -\left(\, \frac{R_v}{r_+} \,\right)^{2d_U-2} \,\right]  }
 \left[\, 1 - \left(\, 2d_u-1\,\right)\,\left(\, \frac{R_v}{r_+} \,\right)^{2d_U-2}\,\right]
\label{tbh}
\end{equation}

by comparing (\ref{tbh} with (\ref{rextr}), we see that

\begin{equation}
 T_{d_U}\left(\, r_+=r_{extr.}\,\right)=0
\end{equation}

As it was expected, the extremal black hole has vanishing Hawking temperature. The second zero-temperature 
configuration is asymptotically approached when $r_+\to\infty$. Thus,
the Hawking temperature increases up a finite maximum value, for $r_{max}> r_{extr}$,
and then drops down to zero as $r_+\to r_{extr.}$\\

It is interesting to consider the temperature in the two  ``phases'' of the model :\\
i) \textit{weak-coupling} phase, where $\lambda<<1$, $R_v<< r_+$;
$T_{d_U} $ takes the standard form

\begin{equation}
 T_{d_U}\simeq T_H=\frac{1}{4\pi r_H}
\end{equation}
 
ii ) \textit{strong-coupling} phase, where  $\lambda>>1$, $R_v>> r_+$;  $T_{d_U}$ turns into

\begin{equation}
 T_{d_U}\simeq \frac{2d_U-1}{4\pi r_H} \label{unt}
\end{equation}

Eq.(\ref{unt}) has the same form as the Hawking temperature for 
Schwarzschild black hole in $D$ spacetime dimensions

\begin{equation}
 T_{D}\simeq \frac{D-3}{4\pi r_H} \label{td}
\end{equation}

It is important to remark that beyond the formal analogy, there is 
a substantial difference between $T_{d_U} $ and $T_{D} $: the topological dimension
$D$ is an integer number while the scaling dimension $d_U$ is a \textit{real} number.
Thus, in the strong coupling phase, the event horizon behaves like
\textit{fractal surface} of \textit{spectral dimension} $d_H=2d_U$.
 Let us elaborate this picture by investigating the Area Law.\\
 We start from the first law of black hole thermodynamics

\begin{equation}
 dM = T_{d_U} dS \label{firstlaw}
\end{equation}
 
where,  $dM= dr_+ \left(\, \partial M/\partial r_+\,\right)$.  Equation (\ref{firstlaw}) 
describes a transformation between two states characterized by a different radius
of the event horizon. This transformation is a ``path'' in the $(M\ ,r_+)$ plane along
a $d_U=const.$ trajectory.

\begin{equation}
dS= \frac{2\pi\, r_+}{\left[\, 1 -  \left(\, \frac{R_v}{r_+} \,\right)^{2d_U-2} \,\right]}\, dr_+
\label{ds}
\end{equation}
 
 In the weak-coupling phase, unvector contributions can be neglected and Equation (\ref{ds}) takes the standard 
 form 
 
 \begin{equation}
dS\simeq 2\pi\, r_+\, dr_+
\label{dsweak}
\end{equation}

which gives after integration the celebrated area-entropy law
\begin{equation}
 S= \pi\, r_+^2 =\frac{1}{4G_N}\,  A_+ 
 \label{1/4}
\end{equation}

In the strong-coupling-phase black hole evolution is different. The key-point is that the final
configuration can be, at most, an extremal black hole, but nothing smaller than that.
Actually, this configuration is asymptotically approached, as it is $T_{d_U}\to 0$ and
smaller and smaller amount of mass is evaporated away. 
Thus, to compute the entropy (change) from Equation (\ref{ds}) the lower integration limit
cannot be smaller than $r_{extr.}$.  In this phase, we find

\begin{equation}
dS\simeq \frac{2\pi}{R_v^{2d_U-2} }\, r_+^{2d_U-1} dr_+
\label{dsstrong}
\end{equation}

and
\begin{equation}
S= \frac{\pi\,R_v^2 }{d_U }\, \left[\, \left(\, \frac{r_+}{R_v}\,\right)^{2d_U}-
\left(\, \frac{r_{extr.}}{R_v}\,\right)^{2d_U}\,\right]
\label{sstrong}
\end{equation}

\section{Conclusions}
We have demonstrated that vector unparticles can modify the Schwarzschild metric for uncharged, unrotating matter, 
creating a Riessner-Nordstr\"om class of solution.  The majority of expected characteristics of the resulting 
black hole -- double horizon, extremality conditions and vanishing temperature, {\it etc...} --  are commensurate 
with the classical case, although we have shown that in the ungravity case there is a minimum (non-zero) inner 
horizon radius, $r_- > R_v$.   The small difference in inner horizon size between the standard RN black hole and 
un-RN solutions suggests there might be deeper discrepancies in the underlying characteristics.  A future path of 
inquiry might be to investigate the influence of unparticles on the physics of the Cauchy horizon 
\cite{Matzner:1979zz,Poisson:1990eh}, which for RN black holes are generally unstable.

The fractal nature of the outer horizon is similar to that obtained for the black holes in \cite{Gaete:2010sp}, 
furthering the notion that unparticles can increase the effective dimensionality of spacetime by a (non-integer) 
number of dimensions.  A greater understanding of the thermodynamics and decay modes of such black holes can 
potentially yield observationally-distinct signatures in current or future experiments.  Lower mass limits on 
primordial un-vector black holes have been previously obtained \cite{Mureika:2009ii}, thus if such objects exist in 
the Universe their evaporation remnants will be visible in this era.

\vskip 1cm
\noindent{\bf Acknowledgements}\\
JRM is supported by the Research Corporation For Science Advancement.

\end{document}